\documentclass[aip,pop,10pt,twocolumn]{revtex4-1}

\usepackage{graphicx}%
\usepackage{dcolumn}%
\usepackage{bm}%
\usepackage{times}%
\usepackage{color}%

\begin{document}

\title{\Large Fire-hose instability of inhomogeneous plasma flows with heat fluxes}

\author{E. S. Uchava}%
\affiliation{\nolinebreak Faculty of Exact and Natural Sciences, Tbilisi State
University, Tbilisi, 0179, Georgia \looseness=-1}%
\affiliation{Evgeni Kharadze Georgian National Astrophysical Observatory, Abastumani, 0301, Georgia \looseness=-1}%
\affiliation{Nodia Institute of Geophysics, Tbilisi State University, Tbilisi,
0171, Georgia \looseness=-1}

\author{A. G. Tevzadze}\email[]{Corresponding author: aleko@tevza.org}%
\affiliation{\nolinebreak Kutaisi International University, Kutaisi University Campus, Kutaisi, 4600, Georgia \looseness=-1}%
\affiliation{Evgeni Kharadze Georgian National Astrophysical Observatory, Abastumani, 0301, Georgia \looseness=-1}%

\author{B. M. Shergelashvili}%
\affiliation{Space Research Institute, Austrian Academy of Sciences, 8042, Graz, Austria \looseness=-1}%
\affiliation{Centre for Computational Helio Studies, Faculty of Natural Sciences
and Medicine, Ilia State University, 0162, Georgia \looseness=-1}%
\affiliation{Evgeni Kharadze Georgian National Astrophysical Observatory, Abastumani, 0301, Georgia \looseness=-1}%

\author{N. S. Dzhalilov}%
\affiliation{Shamakhy Astrophysical Observatory of the ANAS, Baku AZ-1000,
Azerbaijan \looseness=-1}

\author{S. Poedts}%
\affiliation{Center for Mathematical Plasma Astrophysics,
KU Leuven, B-3001, Leuven, Belgium \looseness=-1}%
\affiliation{Institute of Physics, University of Maria Curie-Sk{\l}odowska,
PL-20-031 Lublin, Poland \looseness=-1}%

\begin{abstract}

We study the effects of heat flows and velocity shear on the parallel
firehose instability in weakly collisional plasma flow. For this purpose we
apply an anisotropic 16-moment MHD fluid closure model that takes into
account the pressure and temperature anisotropy, as well as the effect of
anisotropic heat flux. The linear stability analysis of the firehose modes
is carried out in the incompressible limit, where the MHD flow is parallel
to the background magnetic field, while the velocity is sheared in the
direction transverse to the flow direction. It seems that an increase of the
velocity shear parameter leads to higher growth rates of the firehose
instability. The increase of the instability growth rate is most profound
for perturbations with oblique wave-numbers $k_{\perp}/k_{\parallel} < 1$.
{\color{black} Combined action of the velocity shear and heat fluxes
introduce} an asymmetry of the instability growth in the shear plane:
perturbations with wave-vectors with a component in the direction of the
velocity shear grow significantly stronger as compared to those with
components in the opposite direction. We discuss the implications of the
presented study on the observable features of the solar wind and possible
measurements of local parameters of the solar wind based on the stability
constraints set by the firehose instability.
\end{abstract}

\maketitle

\section{Introduction}

It is well known that in low density, ionized, rarefied flows, magnetic field
effects can dominate over the particle collision effects. Such systems can
exhibit different values of temperature and pressure when measured along and
normal to the direction of the background magnetic field. In such flows the
anisotropic effect of the gyration of the charged particles around magnetic field
lines dominates over the isotropic particle collision process leading to a
plasma with anisotropic thermodynamic properties.

Anisotropic ionized flows are prone to a number of kinetic instabilities that
tap energy from the magnetic anisotropy and grow due to various destabilization
mechanisms. Flows where the thermal energy dominates over the magnetic energy
are prone to the firehose
instability\cite{Longmire56,Vedenov58,Chandra58,Parker58,Vedenov61}. When the
pressure parallel to the magnetic field is sufficiently higher than the
perpendicular pressure, transverse kink perturbations of the magnetic field
can become unstable and grow exponentially in time. Also, it has been shown that
in viscous dissipation, anisotropy plays a significant role in the heating of the
solar coronal plasma shear flows \cite{shergelashvili06}, even when other effects are abandoned.

The firehose instability has gained much attention
recenty\cite{Sarfraz2017,Astfalk2016,dosSantos2014,Lazar2013,Schlickeiser2010,Lazar2009,Rosin2011,Hau2007,Wang2003},
since it is believed to be one of the primary kinetic instabilities affecting
the solar and stellar wind dynamics. The importance of the firehose
instability in various astrophysical situations is also
recognized\cite{Rosin2011}. Still, much of the observational evidence about
the development of the instability comes from the measurements of pressure
anisotropies in solar wind
fluctuations\cite{Chen2016,Ibscher2014,Matteini2013,Schlickeiser2011,Bale2009}.

It has been shown that the nonlinear development of the microscopic firehose
instability can affect the large-scale dynamics of astrophysical
plasmas\cite{Schekochihin2008}. A more detailed analysis of the instability
development at nonlinear amplitudes can be conducted by means of numerical
simulations\cite{Seough2015,Rosin2011}. Moreover, direct numerical
simulations of the classical firehose instability using a hybrid-kinetic
numerical approximation confirm the predictions of the standard linear
theory on the exponentially growing instability \cite{Kunz2014}.
In the isotropic limit, the plasma non-equilibrium thermalization manifested by
periodic or rapid aperiodic variations of the system entropy in time, leads to
strong coupling of different MHD wave modes (described in a similar mathematical
framework as for waves in shear flows) and also to development of parametric
instabilities \cite{shergelashvili07}.

The theoretical framework for many of the advances in the analysis of
magnetically anisotropic flows using a fluid description, is adopting the
Chew-Goldberger-Low (CGL) approximation\cite{CGL}. This limit exploits the
simplicity of a fluid description by using a closure model that leads to two
different equations of state, viz.\ parallel and perpendicular to the magnetic
field. Thus, the CGL limit is often referred to as a double adiabatic
anisotropic MHD description.

The purpose of the present paper is to study the properties of the firehose
instability in weakly magnetized anisotropic shear flows with heat fluxes.
Waves and instabilities of anisotropic MHD shear flows have been studied in
the framework of the non-modal approach in the CGL
limit\cite{Chagelishvili1997}. The CGL closure model neglects heat fluxes as
low frequency phenomena and uses the double adiabatic approximation for the
analysis of higher frequency modes. On the other hand, the velocity shear of
the flow is acting on the shearing time-scales: these time-scales are usually
much longer than the short time scales described within the CGL limit and are
comparable to the time scales of thermal processes. Therefore, we here employ
a so-called 16-moment anisotropic MHD model that can be effectively used to
analyze low frequency phenomena in such flows\cite{Oraevskii68,Dzhalilov08}.
This approximation enables us to study anisotropic plasmas with heat fluxes
using the fluid description.
{\color{black} The method is an extension of the 13-moment Grad
method\cite{Grad1949} for the case, taking into account higher order
dispersive effects. These include two anisotropic heat flux vectors that
describe transport of parallel and perpendicular thermal energies and
viscosity tensor that describes collision-free viscosity effect. Viscous
contribution can be often neglected in low frequency limit thus leading to the
fluid model that represents apparent generalization of CGL-MHD model with
addition of anisotropic heat fluxes.}
Indeed, the 16-moment formalism has already proven to be successful in
analyzing waves and instabilities in weakly collisional
media\cite{Kuznetsov2009,Kuznetsov2010,Dzhalilov2011,Uchava14,Ismayilli2018}.

In this paper, we present the stability analysis of the low frequency
incompressible perturbations to the anisotropic MHD shear flows with heat
fluxes. {\color{black} Effects of the velocity shear and heat fluxes on the
dynamics of linear perturbations are characterized by lower frequency compared
to the compressibility effects. Thus, we adopt incompressibility approximation
and study the properties of fire-hose instability when compressibility effects
can not directly affect this instability in the linear limit.}

The physical model of the problem is described in Sec.~2, where the
steady-state flow {\color{black} is described. Section 3 presents the linear}
perturbations and the stability analysis of both static case as well as
sheared flows are described, respectively. The effects of the velocity shear
and anisotropic heat fluxes are summarized in Sec.~3.

\section{Physical Model}

The incompressible anisotropic MHD system can be described in the 16-moment
approximation by using the following simplified fluid
model\cite{Oraevskii68,Dzhalilov08,Uchava14}:
\begin{equation}
{\mathrm{d} \mathbf{V} \over \mathrm{d} t} + {\nabla
P_{\perp} \over \rho} + {\mathbf{B} \times (\nabla \times \mathbf{B}) \over
4 \pi \rho} = {\nabla_{\|} \over \rho} \left(
(P_\perp - P_\|) {{\bf B} \over B^2} \right) ~, \label{MHD1}
\end{equation}
\begin{equation}
{\mathrm{d} \mathbf{B} \over \mathrm{d} t} - \nabla_{\|} \mathbf{V}
= 0 ~, \label{MHD2}
\end{equation}
\begin{equation}
\nabla \cdot \mathbf{B} = 0 ~,~
\end{equation}
and the incompressibility condition:
\begin{equation}
\nabla \cdot\mathbf{V} = 0 ~,
\label{MHD3}
\end{equation}
where the following notations are used for the shortness:
$$
\mathrm{d}/\mathrm{d} t \equiv \partial / \partial t +
\mathbf{V} \cdot \nabla ~,
$$
$$
\nabla_{\|} \equiv {\mathbf{B} \cdot \nabla  \over B} ~.
$$
The equations of state in the parallel and perpendicular to the magnetic field
directions now include the heat fluxes:
\begin{equation}
{\mathrm{d} \over \mathrm{d} t} \left( {P_{\|} B^{2} \over \rho^{3}}
\right) = -{B^2 \over \rho^3} \left( \nabla_{\|}
\left( {S_{\|}\over B} \right) + {2 S_{\perp}\over B^2} \nabla_{\|}  B
\right) ~, \label{CGL1}
\end{equation}
and
\begin{equation}
{\mathrm{d} \over \mathrm{d} t} \left( {P_{\perp} \over \rho B}
\right) = -{B^2 \over \rho} \nabla_{\|} \left( {S_{\perp} \over
B^2} \right) ~. \label{CGL2}
\end{equation}
The CGL MHD equations can be derived by setting the heat flux parameters to
zero ($S_{\perp} = S_{\|}=0$) in Eqs.~(\ref{CGL1}) and (\ref{CGL2}). The full
closure of the system of Eqs.(\ref{MHD1}-\ref{CGL2}) can be accomplished
through the 16-moment closure model that provides two more equations for the
heat fluxes, namely:
\begin{equation}
{\mathrm{d} \over \mathrm{d} t } \left( {S_{\|} B^3 \over \rho^4}
\right) = -{3 P_{\|} B^2 \over \rho^4} \nabla_{\|} \left(
{P_{\|} \over \rho} \right) ~, \label{Heat1}
\end{equation}
and
\begin{equation}
{\mathrm{d} \over \mathrm{d} t} \left( {S_{\perp} \over \rho^2}
\right) = - {P_{\|} \over B \rho^2} \left( \nabla_{\|}
\left( {P_{\perp} \over \rho} \right) + {P_{\perp} \over \rho} {P_{\perp} - P_{\|}
\over P_{\|} B} \nabla_{\|} B \right) ~. \label{Heat2}
\end{equation}

{\color{black} For the sake of completeness we show connection between the
fluid velocity of the flow $\bf V$ and the velocity of individual particles
$\bf u$ in the kinetic regime using particle distribution function $f({\bf
u},{\bf r}, t)$:
$$
{\bf V} = \left( \int{ {\bf u} f {\rm d}^3 {\bf u} } \right) / \left( {\int{f {\rm d}^3 {\bf u} } } \right) ~.
$$
Hence, parallel and perpendicular heat fluxes can be defined using the third
moments as follows:
$$
S_{\parallel} = {m \over 2 B^2} \int{{\rm d}^3 {\bf u} ({\bf u} \cdot {\bf B})^2 u_{\parallel} f } ~,
$$
$$
S_{\perp} = {m \over 2} \int{{\rm d}^3 {\bf u} \left( u^2 - {({\bf u} \cdot {\bf B})^2 \over B^2} \right) u_{\perp} f } ~,
$$
where $m$ and $\bf u$ are mass and velocity of individual particles, while
$u_{\parallel}$ and $u_{\perp}$ stand for the particle velocity components
parallel and perpendicular to the magnetic field, respectively.

}

\subsection{Steady-state equilibrium flow}

We consider a stationary, weakly collisional MHD flow parallel to the uniform
background magnetic field in the $x-$direction, with a velocity profile that
is sheared linearly in the transverse ($y-$)direction:
\begin{equation}
{\bf V_0} = (Ay,0,0) ~,~~ {\bf B_0} = (B_0,0,0) ~,
\end{equation}
where $A$ is the transverse shear parameter of the parallel velocity. Assuming
the similar physical origins of the pressure and heat flux anisotropy in the
rarefied flow, we introduce the pressure anisotropy parameter as follows:
\begin{equation}
\alpha = P_{\perp 0} / P_{\| 0} = S_{\perp 0} / S_{\| 0} ~.
\end{equation}
{\color{black} The pressure and heat flux anisotropy parameters may be
different due to the contribution from different plasma species (e.g., protons
and electrons). Still, we neglect these differences within adopted one fluid
description and use pressure anisotropy parameter as a guiding anisotropy
factor for farther analysis.} Described flow matches the average configuration
of the solar and stellar winds locally, where the flow convexity and turbulent
component can be neglected. Indeed it is known that, while being stable global
configurations, such flows can exhibit a number of micro-instabilities
depending on the magnetic field, anisotropy and heat flux parameters.

{\color{black}
\section{Linear stability analysis}
}

For the purpose of the stability analysis,  we introduce  linear perturbations
of the background  incompressible parallel shear flow embedded in a uniform
magnetic field as follows:
\begin{eqnarray}
{\bf V} &=& {\bf V_0} + {\bf V^\prime} ~,~~~~
{\bf B} = {\bf B_0} + {\bf B^\prime} ~, \nonumber \\
P_\parallel &=& P_{0 \parallel} + P_{\parallel}^\prime ~,~~
P_\perp = P_{0 \perp} + P_{\perp}^\prime ~, \\
S_\parallel &=& S_{0 \parallel} + S_{\parallel}^\prime ~,~~ S_\perp
= S_{0 \perp} + S_{\perp}^\prime ~, \nonumber
\label{Linear}
\end{eqnarray}
where ${\bf V^\prime}\ll {\bf V_0}$, ${\bf B^\prime}\ll {\bf B_0}$, etc.,
{\color{black} $\rho$ is uniform constant density} and
\begin{equation}
C_\|^2\equiv{P_{\|0} / \rho} ~,~~ C_\perp^2\equiv{P_{\perp0} / \rho} ~.
\end{equation}
{\color{black} In present paper we study the effect of background stationary
heat fluxes on the stability of local fire-hose instability. Thus, we
introduce linear perturbations over the nonzero heat fluxes:
$$
S_{0 \parallel} \not= 0 ~,~~~ S_{0 \parallel} \not= 0 ~.
$$
}
Defining the Alfv\'en velocity as
\begin{equation}
V_A = B_0 / \sqrt{4 \pi \rho} ~,
\end{equation}
we may introduce parallel and perpendicular plasma beta parameters as
follows:
\begin{equation}
\beta_{\|} = 4 \pi P_{\| 0} / B_0^2 ~,~~
\beta_{\perp} = 4 \pi P_{\perp 0} / B_0^2  ~.
\end{equation}

Effects of the heat fluxes can be described by the non-dimensional heat flux
parameters \cite{Uchava14}:
\begin{equation}
\gamma_{\|} = \rho S_{\| 0} / P_{\| 0}^2 ~,~~ \gamma_{\perp} =
\rho S_{\perp 0} / P_{\perp 0}^2 ~.
\end{equation}
In the present paper, we introduce parallel and perpendicular heat flux parameters
that also account for the magnetic field of the plasma:
\begin{equation}
q_{\|} = 2 \gamma_{\|} \beta_{\|}^{1/2} ~,~~ q_{\perp} =
2 \gamma_{\perp} \beta_{\perp}^{1/2} ~.
\end{equation}
Hence, upon substituting Eqs.~(\ref{Linear}) into the Eqs.~(\ref{MHD1}-\ref{CGL2})
and neglecting all the nonlinear terms, we can obtain the linear system of partial
differential equations describing the dynamics of perturbations in the
anisotropic MHD flow. This system can be analyzed in the shearing sheet limit,
where spatial Fourier expansion with time dependent wave-numbers can be
employed as follows:
\begin{equation}
\left( \begin{array}{c}
{P_{\|}^\prime({\bf r},t) / P_{\| 0}} \\
{P_{\perp}^\prime({\bf r},t) / P_{\perp 0}} \\
{S_{\|}^\prime({\bf r},t) / P_{\| 0}} \\
{S_{\perp}^\prime({\bf r},t) / P_{\perp 0}} \\
{{\bf V^\prime}({\bf r},t) / V_A} \\
{{\bf B^\prime}({\bf r},t) / B_0}
\end{array} \right)
\propto
\left( \begin{array}{c}
{\rm i} {p_{\|}({\bf k},\tau) } \\
{\rm i} p_{\perp}({\bf k},\tau) \\
s_{\|}({\bf k},\tau) \\
s_{\perp}({\bf k},\tau) \\
{\bf v}({\bf k},\tau) \\
{\rm i} {\bf b}({\bf k},\tau)
\end{array} \right)  \exp \left( {\rm i} {\bf k(\tau)} {{\bf r} \over L} \right)
~. \label{FourierSet}
\end{equation}
Here, $L$ corresponds to the characteristic length-scale of the flow, $\tau$ is the
non-dimensional time variable
\begin{eqnarray}
\tau = t V_A / L ~,
\end{eqnarray}
$k_y(\tau)$ is the non-dimensional shearing wave-number of the perturbation
harmonics:
$$
k_y(\tau) = k_{y0} - R k_x \tau ~,
$$
and $R$ is the normalized velocity shear parameter
$$
R = A L / V_A ~.
$$
{\color{black} Spatial Fourier Harmonics of the physical quantities
($p_\parallel$, $p_\perp$, $s_\parallel$, $s_\perp$, etc.) derived from Eqs.
(\ref{FourierSet}) are in general a complex valued functions. Still, there are
specific phase difference between different quantities on the complex plane.
In uniform compressible flow without heat fluxes the phase of pressure and
velocity perturbations differ by $\pi/2$. This means that perturbations with
real valued velocity harmonics will have purely complex pressure harmonics. To
account for these standard differences we introduce velocity and heat flux
harmonics with real and pressure and magnetic field harmonics with purely
complex factors (see Eq. \ref{FourierSet}).}

In this framework we may derive the system of differential equations governing
the linear dynamics of perturbation harmonics of the incompressible
anisotropic shear flow system:
\begin{eqnarray}
\dot{ v}_x(\tau) &=& R v_y(\tau) - \beta_\perp {k_\perp^2 \over k_x}
p_\perp(\tau) - {\Delta \beta k_x^2 + k^2 \over k_x} b_x(\tau) ~, \label{ODE1}\\
\dot{v}_y(\tau) &=& \beta_{\perp} k_y p_\perp(\tau) +
k_y b_x(\tau) - (1+\Delta \beta) k_x b_y(\tau)~,\\
\label{ODE2}
\dot{ p}_\perp(\tau) &=&  k_x v_x(\tau) + R b_y(\tau) - k_x s_{\perp}(\tau) +
{\rm i} q_{\perp}  \alpha k_x b_x(\tau) ~,\\
\label{ODE3}
\dot{s}_\|(\tau) &=& -{3{\rm i} q_\| \over 2} k_x v_x(\tau) + 6 R v_y(\tau) -
3 \beta_\perp {k_\perp^2  \over k_x} p_\perp(\tau)-~~~~~~~~ \nonumber \\
\label{ODE4}
&-& 3 {2\Delta \beta k_x^2 + k^2 \over k_x} b_x(\tau) -
{3 {\rm i} q_\| \over 2} R b_y(\tau) ~,\\
\dot{ s}_\perp(\tau) &=& \beta_\| k_x p_\perp(\tau) + \Delta \beta k_x b_x(\tau) ~,\\
\label{ODE5}
\dot{b}_x(\tau)&=& R b_y(\tau) + k_x v_x(\tau) ~,\\
\label{ODE6}
\dot{b}_y(\tau) &=&  k_x v_y(\tau) ~,\\
0 &=& k_x b_x(\tau) + k_y b_y(\tau) + k_z b_z(\tau) ~.\label{ODE7}
\end{eqnarray}
where
\begin{eqnarray}
k_\perp^2 &=& k_y^2 + k_z^2 ~, \nonumber \\ k^2 &=& k_x^2 + k_\perp^2 ~, \\
\Delta \beta &\equiv& \beta_\perp - \beta_\| \nonumber ~,
\end{eqnarray}
and the dot denotes the temporal derivative, e.g.\ $\dot{\psi}(\tau) \equiv {\rm d}
\psi / {\rm d} \tau $.
{\color{black} Despite the fact that we have normalized perturbation harmonics
to remove complex parameters due to intrinsic phase differences, Eqs.
(\ref{ODE1}-\ref{ODE7}) still explicitly include complex parameters. Complex
parameters are due to heat fluxes $q_\perp$ and $q_\parallel$.}
{\color{black} Eq. (21) indicates that for non-trivial solutions
($P_{\perp}=S_{\perp}=v_x=b_x=b_y=0$) pressure, velocity, magnetic field and
heat flux perturbations can not be simultaneously real valued variables.}
{\color{black} Thus, heat fluxes to the anisotropic flow brake intrinsic phase
difference between pressure and velocity harmonics and introduce non-removable
phase differences between the harmonics of different physical quantities. }
{\color{black} Note, that incompressibility condition (Eq. 4) is already
applied in the Eqs. ({\ref{ODE1}-\ref{ODE7}}). Here we have chosen
perpendicular pressure perturbation component $p_{\perp}$ for further
analysis. Parallel pressure perturbation can be calculated from the condition
${\rm d}/{\rm d} \tau ({\bf k} \cdot {\bf v}) = 0$. Interestingly, non of the
Eqs. (19-21) and (23-26) contain parallel heat flux perturbation
($s_{\parallel}$) or parallel heat flux parameter ($q_{\perp}$). Thus, under
considered approximation, the ODE system ({\ref{ODE1}-\ref{ODE7}}) can be
split in two parts: parallel heat flux perturbations are driven by other
perturbations (see Eq. (22)). This choice defines that only perpendicular heat
flux parameter appears in further analysis. Still, we note that in present
case parallel and perpendicular heat flux parameters are not independent
parameters and are linked through Eqs. (10) and (16). }

\begin{figure}[t]
\begin{center}
\includegraphics[width = 0.9 \columnwidth]{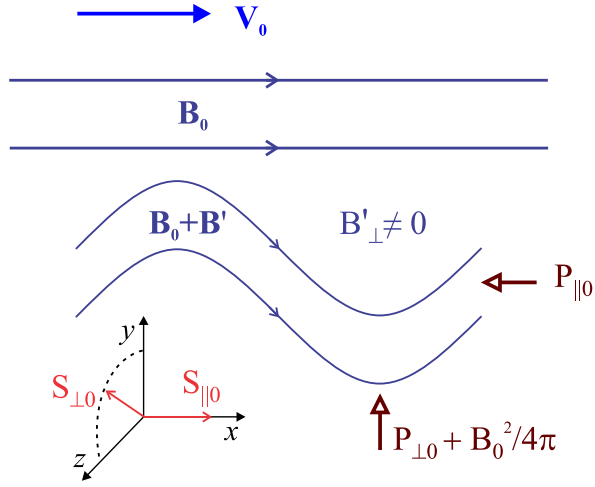}
\caption{Sketch illustration of the firehose instability in the flow parallel
to the magnetic field. Instability occurs when the transverse effective pressure
acting on the perturbed magnetic tube is less than the parallel pressure:
$P_{\perp 0} + B_0^2 / 4 \pi < P_{||0}$ and $c_F^2 <0$. In this case, transverse
Alfv\'enic deformation of the magnetic field lines grow in time exponentially.
{\color{black} Direction of axis adopted for the description of the flow
is shown in the left bottom corner. Arrows on the axis indicate direction of the
equilibrium heat fluxes, subject to linear perturbations.}
 }
\label{sketch}
\end{center}
\end{figure}

\subsection{\color{black} Firehose instability in static case}

In the zero-shear limit ($R=0$), $k_y$ and thus also all coefficients in the
differential equations become time-independent. Hence, we may use a spectral expansion of the linear
perturbations in time $\propto \exp({\rm i} \omega \tau)$ and obtain
the  dispersion equation of the incompressible, anisotropic MHD system, which reads as follows:
\begin{equation}
\left[ \omega^2 - (1+\Delta \beta) k_x^2 \right] D_0 = 0 ~,
\label{Disp0}
\end{equation}
where
\begin{eqnarray}
D_0 \equiv \omega^4 - (1+\beta_\perp)k^2 \omega^2 +
\alpha \beta_\perp q_\perp k_x k_\perp^2 \omega +  \hskip 1.5 cm \\
~~~ \left[ (1+\Delta \beta) k_x^2 + (1+\beta_\perp-\alpha^2 \beta_\|)
k_\perp^2 \right] \beta_\| k_x^2 ~. \nonumber
\end{eqnarray}
The first obvious solution of this dispersion equation is given by the
firehose mode:
\begin{equation}
\omega_{\rm F \pm} = \pm c_F k_x  ~,
\label{Firehose0}
\end{equation}
where
\begin{equation}
c_F^2  \equiv 1 + \Delta \beta~,
\end{equation}
stands for the square of the characteristic speed of the firehose mode, when
positive. It seems that kink deformations of the parallel magnetic structures
that are described by the firehose mode do not feel the effect of heat fluxes
($q_\|$, $q_\perp$). Hence the linear stability criterion for firehose
instability ($c_F^2<0$) can be set by the balance of the parallel and
perpendicular plasma beta {\color{black}
parameters\cite{Parker58,Vedenov61,Evangelias}:}
\begin{equation}
\beta_\| > 1+\beta_\perp ~.
\end{equation}
The growth of linear perturbations is described by $\omega_{\rm F+}$ or
$\omega_{\rm F-}$, depending on whether the streamwise wave-number $k_x$ is
positive or negative, respectively.

The mechanism of the classical microscopic anisotropic MHD instability can be
illustrated using a simple sketch shown in Fig.~(\ref{sketch}). Transverse
perturbations of the magnetic field ($B^\prime_\perp \not=0$) can be shown as
kink perturbations of the parallel magnetic structure. The response to this
perturbation consists from a combined action of the parallel and perpendicular
pressure and the magnetic field. If the perpendicular pressure is lower than
the parallel pressure to the extent that even the magnetic field can not
compensate for the kink deformation, the perturbation will grow and the
instability will be developed. Interestingly, the heat fluxes ($S_\|$,
$S_\perp$) do not affect this process. At least not in the static/uniform flow
limit.

{\color{black} Another set of modes in the Eq. (\ref{Disp0}) are described by
$D_0=0$ solutions: these are incompressible slow modes modified by the thermal
effects. In the static case these modes are decoupled with the firehose
instability and does not explicitly affect their dynamics.}

\begin{figure*}[t]
\begin{center}
\includegraphics[width = 0.9 \textwidth]{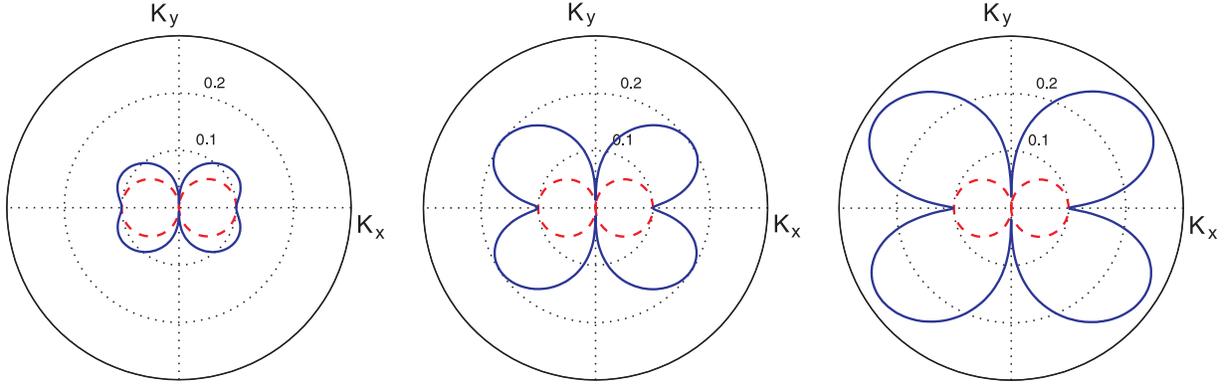}
\caption{
The growth rate of the firehose instability as function of the angle
$\phi$ of the wave vector with respect to the direction of the magnetic field
{\color{black} in the shearing plane: $k_x = k_0 \cos(\phi)$ and $k_y = k_0 \sin(\phi)$ with $k_0=1$.}
Here, $\alpha=0.5$, $\beta_\|=2$, $q_\perp=0.2$, $k_z=1$, $c_F^2=-0.01$ and
$R = 0.05, 0.2, 0.4$ (left to right). The horizontal axis corresponds to the
direction parallel and the vertical axis to the direction perpendicular to the magnetic field,
respectively. The red dashed line shows the instability growth rate for static case
($R=0$), while the solid blue line shows the velocity shear modification.
It seems that an increase of the shear parameter leads to stronger instabilities
for perturbations with oblique wave-numbers: $\phi\sim \pi/4$.}
\label{PolarR}
\end{center}
\end{figure*}

\begin{figure*}
\begin{center}
\includegraphics[width = 0.9 \textwidth]{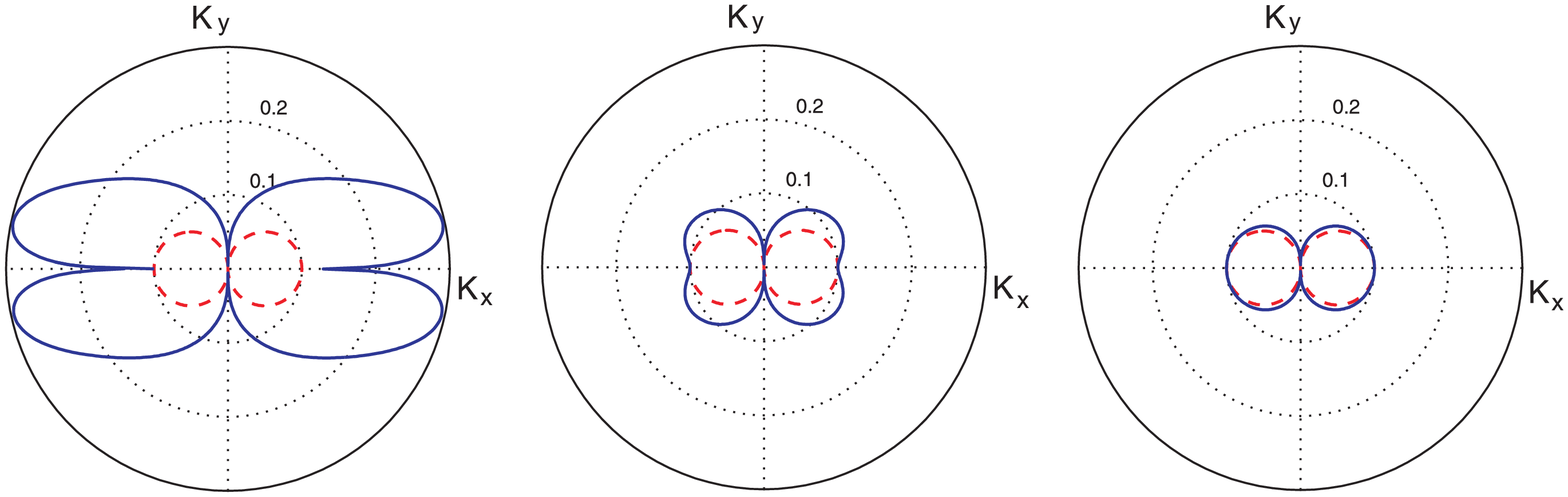}
\caption{
The growth rate of the firehose instability as function of the angle
$\phi$ of the wave vector with respect to the direction of the magnetic field.
Here, $\alpha=0.5$, $\beta_\|=2$, $q_\perp=0.2$, $R=0.05$, $c_F^2=-0.01$  and
$k_z = 0, 1, 2$ (left to right). It seems that the velocity shear effect is the strongest
for vertically uniform perturbations ($k_z=0$), while small scale perturbations
with $k_z >1$ remain largely insensitive to the velocity shear effects.
} \label{PolarKz}
\end{center}
\end{figure*}

\begin{figure*}
\begin{center}
\includegraphics[width = 0.9 \textwidth]{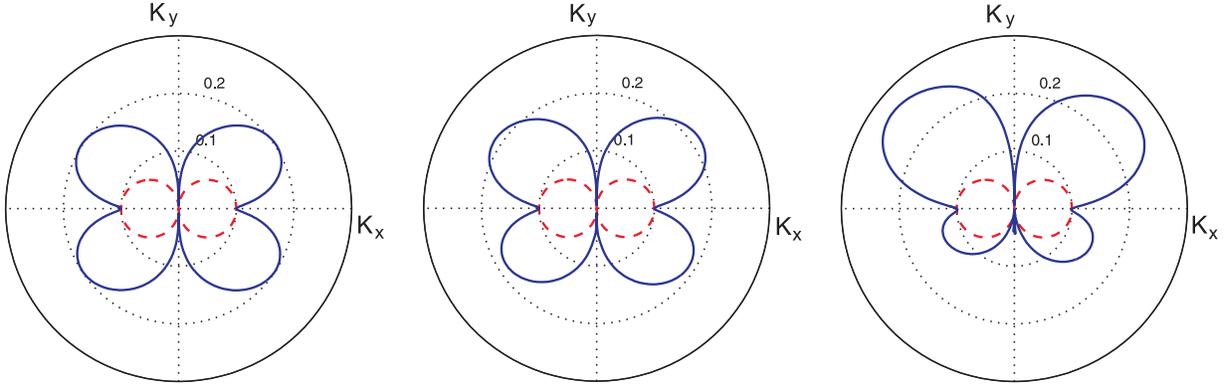}
\caption{
The growth rate of the firehose instability as function of the angle
$\phi$ of the wave vector with respect to the direction of the magnetic field.
Here, $\alpha=0.5$, $\beta_\|=2$,  $k_z=1$, $c_F^2=-0.01$, $R=0.2$ and
$q_\perp=0.2, 2, 10$ (left to right). It seems that an increase of the heat flux
parameter ($q_\perp$) leads to an asymmetry of the instability growth in
shear flows.} \label{PolarQ}
\end{center}
\end{figure*}

\begin{figure*}
\begin{center}
\includegraphics[width = 0.9 \textwidth]{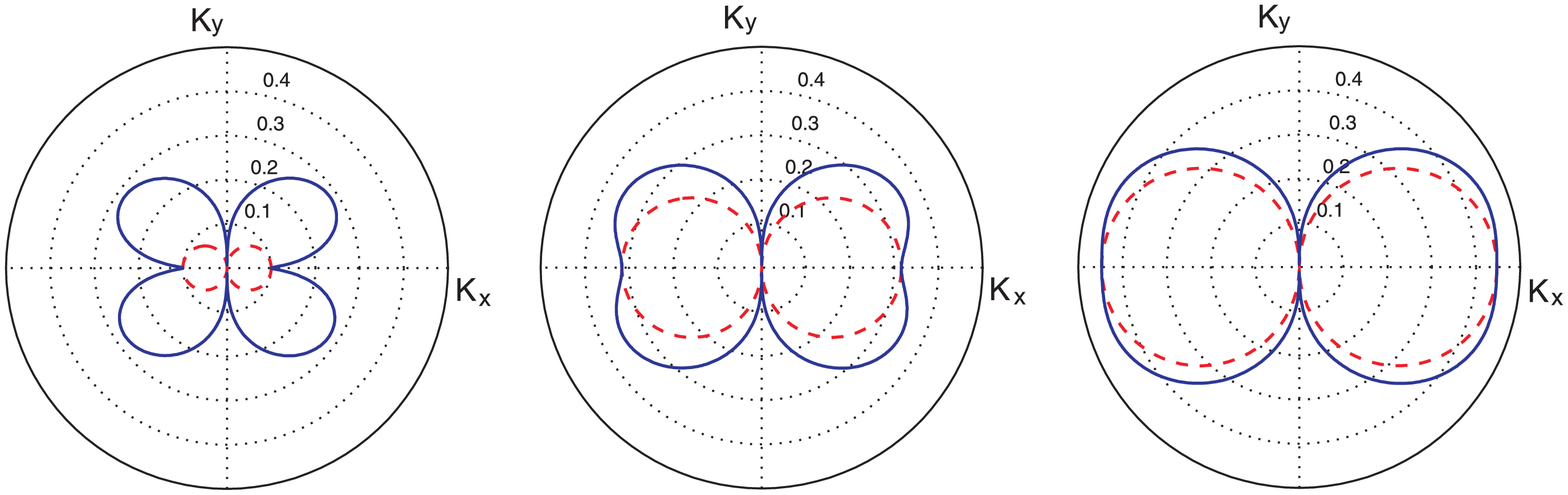}
\caption{
The growth rate of the firehose instability as function of the angle
$\phi$ of the wave vector with respect to the direction of the magnetic field.
Here, $\alpha=0.5$, $\beta_\|=2$, $k_z=1$, $R=0.4$ and
$c_F^2= -0.01, -0.1, -0.2$ (left to right). The velocity shear effects are most
profound for weakly unstable perturbations. {\color{black} Otherwise the
growth rate of firehose instability is lower than the velocity shear parameter
$\sigma^2 < R^2$ and analytic results derived within the adiabatic approximation
are not valid.}
} \label{PolarC}
\end{center}
\end{figure*}

\subsection{Firehose instability in non-uniform flows}

The dynamic behaviour of linear perturbations in shear flows ($R\not=0$) are
described by Eqs.~(\ref{ODE1}-\ref{ODE7}). {\color{black} Employing the low
frequency limit we use the following approximation for time variation of
perturbation harmonics:}
\begin{equation}
\dot{\psi}({\bf k},\tau) \approx - {\rm i} \omega(\tau) \psi({\bf k},\tau) ~.
\end{equation}
{\color{black} This limit is somewhat similar to the adiabatic limit, when
temporal variations of the background are slow enough ($R \ll 1$) to introduce
parameter similar to the spectral frequency $\omega(\tau)$. Such method is
widely used in the adiabatic approximation, where frequency varies slower then
the oscillations it describes.
\begin{equation}
| \dot \omega(\tau) | \ll \omega^2(\tau) ~.
\label{Adiabat}
\end{equation}
In our case the source of time variation of frequency (or growth rate in case
of instability) is the velocity shear parameter leading to the wave-number
variation in the shearing sheet limit:
\begin{equation}
\omega(\tau) = \omega(k_x,k_y(\tau),k_z) ~.
\end{equation}
Using this assumption
\begin{equation}
\dot \omega(\tau) = {\partial \omega  \over \partial k_y}  {\partial k_y(\tau) \over \partial \tau} ~.
\label{OmegaDot}
\end{equation}
and substituting Eq. (\ref{OmegaDot}) into the condition (\ref{Adiabat}) we
derive the following:
\begin{equation}
|\omega^2| \gg |R k_x {\partial \omega \over \partial k_y} |
\end{equation}
}

In this limit, we can derive the adiabatic dispersion equation of the shear
flow system and obtain:
\begin{equation}
\left[ \omega^2 - (1+\Delta \beta) k_x^2 \right] D_0 +
{\rm i} R k_x k_y D_1 = 0 ~, \label{DispR}
\end{equation}
where
\begin{equation}
D_1 \equiv (1+\beta_\perp) \omega^2 - \alpha \beta_\perp q_\perp k_x \omega +
(\beta_\perp^2 -\beta_\| \beta_\perp -\beta_\|) k_x^2~,
\end{equation}
shows the modification to the standard dispersion equation.
The implications of the velocity shear effects on the firehose instability set
by Eq.~(38) are obvious. Unlike the static/uniform flow result, the firehose
solutions now do depend on the velocity shear $R$ as well as on the heat flux
parameter $q_\perp$. {\color{black} In addition, the velocity shear effect may
lead to asymmetry of the growth rates in the shearing plane (see last term of
the Eq. (38)).}

{\color{black} Modes described by Eq. (38) can be coupled due to the velocity
shear of the flow. Low frequency resonance can occur between unstable firehose
and MHD wave modes. Similar effect was studied in CGL
limit\cite{Chagelishvili1997} and it was shown that moderate shear parameter
is needed for such resonant coupling to occur. }

To analyze the effect of shear flow on the classical firehose mode we
introduce the deviation from the static solution (\ref{Firehose0}) as follows:
\begin{equation}
\omega = \omega_{\rm F \pm} + R \omega_{1\pm} ~.
\label{omega1}
\end{equation}
{\color{black} This limit can be especially useful in the low shear limit,
when $R<1$, i.e., shearing time-scale is much longer than the Alfv\'en wave
time-scales in the flow. Using Eq. (\ref{omega1}) in the condition (37) and
recalling that fire-hose mode frequency $\omega_{\rm F\pm}$ does not depend on
the $k_y$ we get the applicability condition of our adiabatic approximation:
\begin{equation}
|\omega^2| \gg R^2 \left| k_x {\partial \omega_{1\pm} \over \partial k_y } \right| ~.
\label{adiabat2}
\end{equation}
}

Substituting Eq.~(\ref{omega1}) into the Eq.~(\ref{DispR}) and neglecting
terms of order higher then $R^2$ in the low shear limit ($R<1$), we may derive the
second-order dispersion equation with respect to  $\omega_{1\pm}$ as
follows:
\begin{equation}
R A_\pm \omega_{1\pm}^2 + (B_\pm + {\rm i}R C_\pm ) \omega_{1\pm} + {\rm i}
D_\pm = 0 ~,
\label{DispF}
\end{equation}
where
\begin{eqnarray}
A_{\pm} &=& 4 c_F^2 (c_F^2-\beta_\|) k_x^2 +
(3 \Delta_\pm - 2(1+2\beta_\perp ) ) k_\perp^2 ~,
\\
B_{\pm} &=& 2 \omega_{F\pm} \Delta_{\pm} k_\perp^2 ~,
\\
C_{\pm} &=& (1 + 2 \beta_\perp - 2 \Delta_\pm ) k_x k_y ~,
\\
D_{\pm} &=& - \omega_{F\pm} \Delta_{\pm} k_x k_y ~,
\end{eqnarray}
and
\begin{eqnarray}
\Delta_{\pm} = \pm c_F \alpha \beta_\perp q_\perp +
(1-c_F^2)(1 + 2 \beta_\perp) - c_F^2 ~.
\end{eqnarray}
The solution of the Eq.~(\ref{DispR}) should match the standard firehose
solution in the zero shear limit, i.e.\
$$
\omega(R=0) = \omega_{F\pm} ~.
$$
This sets the following convergence requirement on the shear flow correction:
\begin{equation}
\lim_{R\to0} \left( R \omega_{1\pm} \right)  =0 ~.
\label{limit}
\end{equation}
Hence, from the two solutions of the reduced dispersion equation
(\ref{DispF}), we can choose the one obeying the asymptotic condition
(\ref{limit}):
\begin{equation}
\omega_{1\pm} = {B_\pm + {\rm i} R C_{\pm} \over 2 R A_{\pm}}
\left[ -1 +
\left(1 - {4{\rm i}R A_\pm D_\pm \over (B_\pm + {\rm i} R C_{\pm})^2}
\right)^{1 / 2}
\right] ~. \label{Sol0}
\end{equation}
Using the low shear limit ($R<1$) we may separate the real and complex parts
of the solution (\ref{Sol0}) for stable, unstable and neutrally stable
firehose modes:
\begin{eqnarray}
\omega =& \pm c_F k_x + \delta_{1\pm} + {\rm i} \sigma_{1\pm} ~~~~~~~~~~~~~~~~~~~~
&{\rm ~when~} c_{F}^2 > 0 ~, \label{Sol1} \\
\omega =& {\rm 2i} \sigma_A ~~~~~~~~~~~~~~~~~~~~~~~~~~~~~~~~~~~~~~~~~~~~~~~~
&{\rm ~when~} c_{F}^2 = 0 ~, \label{Sol2} \\
\omega =& (\sigma_A + \delta_{2\pm}) + {\rm i} \left( \pm|c_F| k_x + \sigma_{2\pm}
\right) &{\rm ~when~} c_{F}^2 < 0 \label{Sol3} ~,~~~~~
\end{eqnarray}
where
$$
\sigma_A \equiv R {k_x k_y \over 2k_\perp^2},
$$
describes the shear flow induced transient growth of aperiodic perturbations,
while the explicit forms of $\delta_{1\pm}$, $\delta_{2\pm}$, $\sigma_{1\pm}$,
$\sigma_{2\pm}$ are given in the Appendix~A. {\color{black} Applicability of
the results derived in Eqs. (50-52) depend on the validity of the adiabatic
approximation (see Eq. (41)). Results are valid if the growth rate of the
instability is larger than the velocity shear of the flow: $\omega^2 > R^2$. }

\section{Discussion}

In order to show the effects of different physical factors and parameters on
the firehose instability, we illustrate analytic solutions given by Eq.
(\ref{Sol3}).

Fig.~(\ref{PolarR}) shows the comparison of the growth rates of the
instability in a static case to flows with different values of the velocity
shear parameter. It seems that the velocity shear effect is negligible for
strictly parallel {\color{black} in the shear plane ($k_y=0$)} or
perpendicular {\color{black}($k_x=0$)}  perturbations {\color{black}(see also
Eq. (45)).} However, velocity shear induces a significant boost to the
instability growth rates for perturbations with wave-vectors oblique to the
magnetic field.

Fig.~(\ref{PolarKz}) shows the growth rates of the instability in static case
and sheared flows for different values of the vertical wave-number. It seems
that the effect of the velocity shear is most profound on the vertically
uniform perturbations with $k_z=0$. The increase of the instability growth
rates is most profound for nearly parallel perturbations with $k_x/k \lesssim
1$.

Fig.~(\ref{PolarQ}) shows the growth rates of the instability in flows with
different heat flux parameters. Interestingly, {\color{black} combined action
of velocity shear and heat fluxes} introduces an asymmetry of the firehose
instability growth in wave-number space: perturbations having a transverse
component pointing in the direction of the velocity shear ($k_y>0$), are
amplified, while the perturbations with wave vectors pointing in the opposite
direction ($k_y<0$), are somewhat suppressed. Notably weaker asymmetry is
introduced in the streamwise direction: perturbations in the $k_x>0$ area grow
stronger in comparison to those in the $k_x<0$ area.

Finally Fig.~(\ref{PolarC}) shows the growth rates of the instability for
different values of the firehose parameter $c_F^2$. The effects of the
velocity shear are most profound for marginally unstable perturbations, while
the violent firehose processes remain largely insensitive to the velocity
shear modifications.

It seems that low frequency processes in the anisotropic flows that are
marginally unstable to firehose perturbations can be significantly modified by
the velocity shear as well as by heat fluxes of the MHD medium. The parallel
firehose instability acquires a transverse component in shear flows and, hence, is
affected by the heat fluxes. Thus the observable features of the MHD
fluctuations can be significantly modified in high shear regions of
anisotropic flows.

In solar and stellar winds the combined effect of the velocity shear and heat
fluxes can be an additional source of turbulence and anomalous heating in
rarefied magnetized outflows. The firehose instability can lead to enhanced
turbulence in  anisotropic MHD flows. Thus, sustained fluctuations of the
solar wind can occur in areas of phase space where no instabilities are
present. Therefore, observations of the fluctuations in the solar wind can be
a powerful tool for analyzing the physical conditions in the wind using the
stability considerations for local perturbations. Indeed, the firehose
instability modified by the effects of velocity shear and heat fluxes can draw
the energy of the background flow into turbulent fluctuations and, ultimately,
into heat via dissipative effects which can be included in the solar/stellar
models in terms of statistically proven macroscopic wave heating and pressure
gradient quantities\cite{shergelashvili12}. It is also interesting to
elaborate on the role of the firehose instability in the dynamics of the
smaller scale flows like solar coronal jets, for which quasi-oscillatory
precursors in the mean intensity variations recently have been
observed\cite{Bagashvili18} that pretend to be triggers of the instability
processes occurring in coronal bright points.

Thus, the analytic solutions derived in the present paper can be used to study the
features of the solar wind fluctuations by deducing the physical conditions in
the outflow, such as the parameters of the heat fluxes and the azimuthal velocity
shear of the radial outflow.

\section*{Acknowledgements}
The work was supported by Shota Rustaveli National Science Foundation grants
DI-2016-52 and FR17\_609.

\section*{DATA AVAILABILITY}

The data that supports the findings of this study are available within the
article and its appendix.

\appendix \section*{Appendix: Firehose solution in non-uniform flows}
\setcounter{section}{1} \setcounter{equation}{0}

The modification of the firehose solutions in non-uniform flows can be
described in low shear limit ($R<1$) using Eqs. (\ref{Sol1}-\ref{Sol3}). In
these equations we have introduced number of notations for the shortness of
the presentation in the main text of the article. Shear flow modification in
the stable ($c_F^2>0$) flow configuration to the real:
\begin{equation} \delta_{1\pm}={k_\perp(t)^2\over
8 c_F k_x A_\pm \Delta_\pm}\left\{M_{1\|}k_x^2+{|k_x\Delta_\pm|\over
k_x\Delta_\pm}\sigma_A^2[2\beta_\perp+1-2\Delta_\pm]^2\right\},
\end{equation}
and imaginary parts of the frequency $\omega$:
\begin{equation}
\sigma_{1\pm}=\sigma_A{P_{1\|}k_x^2+P_{1\perp}k_\perp(t)^2\over
2A_\pm} ~,
\end{equation}
{\color{black} introducing}
\begin{equation}
\eta \equiv  \alpha |c_F| \beta_\perp q_\perp ~,
\end{equation}
and
\begin{eqnarray}
M_{1\|} &=& 8c_F^2\Delta_\pm^2\left(-1+{|k_x\Delta_\pm|\over
k_x\Delta_\pm}\right)~,
\\
P_{1\|} &=& 4[(c_F^2-1)(c_F^2-\beta_\|+2)-\beta_\perp+1] ~,
\\
%Q_{1\|} &=& 0 ~,
%\\
P_{1\perp} &=& \Delta_\pm+\left(1+{|k_x\Delta_\pm|\over
k_x\Delta_\pm}\right)[\pm\eta-2c_F^2(1+\beta_\perp)] ~.
%\\
%Q_{1\perp} &=& ... ~,
\end{eqnarray}
Shear flow modification in the unstable flow configuration ($c_F^2<0$) to the
real:
\begin{eqnarray}
\delta_{2\pm}&=&-\sigma_A \alpha\beta_\perp q_\perp
k_\perp(t)^2\times
\\
&&\times{(M_{2\|}+\eta^2N_{2\|})k_x^2+(M_{2\perp}+\eta^2N_{2\perp})k_\perp(t)^2\over
T_2}\nonumber~,
\end{eqnarray}
and imaginary parts of the frequency $\omega$:
\begin{eqnarray}
\sigma_{2\pm}&=&\mp{\sigma_A k_\perp(t)^2\over 2
|c_F|}\times
\\
&&\times{(P_{2\|}+\eta^2Q_{2\|})k_x^2+(P_{2\perp}+\eta^2Q_{2\perp})k_\perp(t)^2\over
T_2}\nonumber~.
\end{eqnarray}
where
\begin{eqnarray}
T_2&=&[\eta^2-(2c_F^2(1+\beta_\perp)-(1+2\beta_\perp))^2]\{[4c_F^2\times
\\
&& \times(c_F^2-\beta_\|)k_x^2-(6c_F^2(1+\beta_\perp)-\nonumber
\\
&&-(1+2\beta_\perp))k_\perp(t)^2]^2+9\eta^2k_\perp(t)^4\}\nonumber~,
\\
M_{2\|} &=&4c_F^2(c_F^2-\beta_\|)(-2\beta_\perp(-2c_F^2+1)-
\\
&&  -4\beta_\|+1)\times(2\beta_\perp(2c_F^2+4\beta_\|-5)-4\beta_\|+3)\nonumber~,
\end{eqnarray}
\begin{eqnarray}
N_{2\|} &=& 16c_F^2k_x^2(c_F^2-\beta_\|)~,\nonumber
\\
M_{2\perp} &=& 3[1+2(c_F^2+\beta_\|)(2c_F^2-1)]^2[2(c_F^2-1)(c_F^2+\nonumber
\\
&&+\beta_\|)+1]+[6(c_F^2+\beta_\|)(c_F^2+2\beta_\|-2)+1+2\beta_\perp]\times\nonumber
\\
&& \times[-2\beta_\perp(-2c_F^2+1)-4\beta_\|+1][2\beta_\perp\times\nonumber
\\
&& \times(2c_F^2+4\beta_\|-5)-4\beta_\|+3]\nonumber~,
\\
N_{2\perp}&=&48(c_F^2+\beta_\|)(\beta_\|-1)+4+8\beta_\perp ~,\nonumber
\\
P_{2\|} &=&4(c_F^2-\beta_\|)c_F^2[1+2(c_F^2+\beta_\|)(2c_F^2-1)]^2\times\nonumber
\\
&& \times[2(c_F^2-1)(c_F^2+\beta_\|)+1]\nonumber~,
\\
Q_{2\|} &=& 32c_F^4(\beta_\|^2-c_F^4)~,\nonumber
\\
P_{2\perp} &=& [6(c_F^2+\beta_\|)(c_F^2+2\beta_\|-2)+1+2\beta_\perp]\times\nonumber
\\
&&[1+2(c_F^2+\beta_\|)(2c_F^2-1)]^2(2(c_F^2-1)(c_F^2+\beta_\|)+1)\nonumber ~,
\\
Q_{2\perp} &=& 12\eta^2+3([2\beta_\perp(-2c_F^2+1)-4\beta_\|+1]
[2\beta_\perp\times\nonumber
\\
&&\times(c_F^2+2\beta_\|-2)+1+2\beta_\perp](c_F^2+\beta_\|)c_F^2+4\beta_\|-5)-
\nonumber
\\
&&-4\beta_\|+3]-8[6(c_F^2+\beta_\|)(c_F^2+2\beta_\|-2)+1+2\beta_\perp]\times\nonumber
\\ &&\times(c_F^2+\beta_\|)c_F^2~. \nonumber
\end{eqnarray}

\end{document}